\documentclass[prb,twocolumn,showpacs]{revtex4}

\usepackage{amssymb,amsmath}
\usepackage[mathscr]{eucal}

\usepackage{graphicx}

\begin{document}

\title{Transport properties of single atoms.}

\author{
Alexei Bagrets$^{1}$, Nikos Papanikolaou$^{2}$, and Ingrid Mertig$^{1}$ }

\affiliation{
$^1$Martin-Luther-Universit\"at Halle-Wittenberg, Fachbereich Physik, D-06099 Halle, Germany \\
$^2$Institute of Microelectronics, NCSR "Demokritos", GR-15310 Athens, Greece }

\date{\today}

\begin{abstract}
We present a systematic study of the ballistic electron
conductance through $sp$ and $3d$ transition metal atoms attached
to copper and palladium crystalline electrodes. We employ the
{\it ab initio} screened Korringa-Kohn-Rostoker Green's function
method to calculate the electronic structure of nanocontacts while
the ballistic transmission and conductance eigenchannels were
obtained by means of the Kubo approach as formulated by Baranger
and Stone. We demonstrate that the conductance of the systems is
mainly determined by the electronic properties of the atom
bridging the macroscopic leads. We classify the conducting
eigenchannels according to the atomic orbitals of the contact atom
and the irreducible representations of the symmetry point group of the
system that leads to the microscopic understanding of the
conductance. We show that if impurity resonances in the density of
states of the contact atom appear at the Fermi energy,
additional channels of appropriate symmetry could open.
On the other hand the transmission of the existing channels
could be blocked by impurity scattering.

\end{abstract}


\maketitle

\section{Introduction}
State-of-the-art experimental techniques, like
scanning tunneling microscopy (STM)\cite{Ohnishi}
and mechanically controllable break junctions (MCBJ),\cite{Muller,MCBJ}
enable the manipulation of individual atoms and
the fabrication of metallic, atomic-sized contacts.
From a fundamental scientific point of view
these systems allow us to investigate a variety of phenomena
at the ultimate limit of atomic scale and prove basic ideas
of quantum mechanics.
For example, when studying conductance,
the familiar Ohm's law completely breaks down and the fully quantum-mechanical
description should be applied to elucidate
the phenomenon.\cite{Lang,Kobayashi,Cuevas_TB_model,TB_models,Ab-initio,Ab-initio_1}

Our understanding of ballistic transport of atomic-sized conductors
is usually based on the Landauer formula,\cite{Landauer}
\[
 G = \frac{2e^2}{h} \sum_{n=1}^{N} T_n,
\]
where conductance is represented
as a sum of transmission probabilities $T_n$
of individual eigenchannels. When the $T_n$'s are known
many properties of a system can be predicted such as
conductance fluctuations,\cite{cond_fluct} the shot noise,\cite{shot_noise}
dynamical Coulomb blockade\cite{CB} and the supercurrent.\cite{supercurr}
It has been demonstrated in the pioneering
work by Scheer {\it et al.}\cite{Scheer}
that a study of the current-voltage characteristics for the superconducting
atomic-sized contacts allows to obtain transmission probabilities
$T_n$'s for particular atomic configurations
realized in the MCBJ experiments. The $T_n$'s can be determined
by fitting theoretical and experimental $I-V$ curves
and exploiting information contained in the
"subgap structure".\cite{Scheer,Scheer_Nature}
The breakthrough in the interpretation of experimental
results was done by Cuevas {\it et al.}\cite{Cuevas_TB_model}
who suggested a parameterized tight-binding
model for the description of atomic constrictions.
As a central result it has been shown that the number of
conducting modes (eigenchannels) for a single-atomic contact correlates
with the number of valence orbitals of the contact atom
which therefore determines the current flowing through the system.
This result has been confirmed experimentally
by Scheer {\it et al.}\cite{Scheer_Nature}
The analysis of the last conductance plateau corresponding
to single-atom constrictions has revealed that a number
of conducting channels is 1 for Au, 3 for Al and Pb,
and 5 for~Nb.

The main goal of this paper is to clarify
with the help of parameter-free {\it ab initio} calculations
the issue on how the valence states of the central atom
of a single-atom contact determine conducting channels. Our approach is based
on the screened Korringa-Kohn-Rostoker (KKR) Green's function
method\cite{SKKR} supplemented with the
Baranger and Stone\cite{BarStone} formulation of the conductance problem.
Instead of considering a variety of different metallic atomic point contacts
we restrict our calculations to single $sp$
or $3d$ transition metal atoms attached to noble metal (Cu)
or transition metal (Pd) leads. Systems of this type can be realized
experimentally by using alloys to fabricate the point
contacts.\cite{Enomoto,Heemskerk} According to our preliminary study,\cite{Pd_imp}
we await the variation of conductance
depending on the atomic number of the impurity situated
at the central site. The number of conducting
modes is also expected to change with respect to the atomic
number of the impurity, especially because of the existence of the
virtual bound states. We would like to mention, that
the conductance through different type of constrictions
contaminated by impurities has been studied before on the {\it ab initio} level
by Lang\cite{Lang_imp}, Hirose and coworkers\cite{Hirose} and
Palot{\'a}s and coworkers.\cite{Palotas}
In contrast to previous studies, we analyze
the eigenchannel decomposition of the conductance.
Within our approach each channel is classified according
to the irreducible representation of the corresponding point group
of the system as well as by orbital contributions when the channel wave function
is projected on the central atom. That relates
the method presented here with the tight-binding model suggested
by Cuevas {\it et al.}\cite{Cuevas_TB_model}
and gives a microscopic insight into the conductance.

\section{Theoretical approach}

\subsection{Set-up of the problem}

The structures under study consist of two semi-infinite Cu or Pd
fcc (001) leads connected by an atomic cluster with a single
contact atom in the middle of a junction as it is shown in Fig.1.
We consider the contact atom to be either a host atom or different
$sp$ or $3d$ transition metal impurities as will be discussed later.
We assume that the geometry of constrictions shown in Fig.1
catches limiting configurations of atomic contacts produced in the
MCBJ experiments. To realize different contact atoms one could use dilute
alloys as a wire material in the MCBJ,\cite{Enomoto,Heemskerk}
or probe conductance through a single atom with an STM tip.\cite{Gimzewski}
Since in the experiments the measurements are not
performed for the equilibrium atomic configurations rather than
the conductance is recorded dynamically under stretching of
nanocontacts,\cite{MCBJ} we did not optimize the interatomic distances and
kept them the same as in bulk fcc metals Cu and Pd with lattice
constants 6.83 a.u.\ and 7.35 a.u., respectively.

\subsection{Electronic structure calculation of atomic contacts}

 Our calculations are based on density functional theory within
the local density approximation. We employed the non-relativistic version
of the screened Korringa-Kohn-Rostoker
(KKR) Green's function method to calculate the electronic
structure of the systems. Since details of the method
can be found elsewhere,\cite{SKKR_review}
only a brief description is given below.
The parametrization of Vosko, Wilk, and Nusair
\cite{Vosko_Wilk_Nusair} for the exchange and correlation
energy was used. The potentials were
assumed to be spherically symmetric around each atom
(atomic sphere approximation, ASA). However, the full charge
density, rather than its spherically symmetric part, was taken into account.
To achieve well converged results the angular momentum cut-off for the
wavefunctions and the Green's function was chosen to be $l_{\rm{max}}=3$ that
imposed a natural cut-off $2 l_{\rm{max}}=6$ for the
charge density expansion.
In case of the heavy element Pd the scalar
relativistic approximation\cite{SRA} was used.

  In the multiple-scattering KKR approach the one-electron
retarded Green's function is given by the site angular momentum
expansion
\begin{eqnarray}
\nonumber
\lefteqn{G^+({\mathbf R}_n +\ {\mathbf r},
{\bf R}_{n'} +\ {\mathbf r'};E) } \\
\label{Grr}
& = & \delta_{nn'}\sqrt{E} \sum_L R_L^n({\bf r}_<;E)H_L^n({\bf r}_>;E) \\
\nonumber
&  &\  +\ \sum_{LL'}R_L^n({\bf r};E)G_{LL'}^{nn'}(E)R_{L'}^{n'}({\bf r'};E)
\end{eqnarray}
where ${\mathbf r}$, ${\mathbf r'}$ are restricted to the cells
$n$ and $n'$; ${\mathbf r}_<$, ${\mathbf r}_>$ denote one of
the two vectors ${\mathbf r}$ or ${\mathbf r'}$ with the smaller
or the larger absolute value, and local functions
$R_L^n({\bf r};E)$ and $H_L^n({\bf r};E)$ are
the regular and irregular solutions of the
Schr\"odinger equation for the single potential $V_n({\mathbf r})$
of the $n$-th cell in free space. Here the index $L = (l,m)$ stands
for the angular momentum quantum numbers and atomic units are used:
$e = -\sqrt{2}$, $\hbar = 1$, $m = 1/2$.
The structural Green's function $G_{LL'}^{nn'}(E)$  (structure constants)
in Eq.~(\ref{Grr}) is related to the known structure
constants of the appropriately chosen reference system by the
algebraic Dyson equation which includes the difference
$\boldsymbol{\Delta t} =
\delta_{nn'} \delta_{LL'} \Delta t^{n}_{L}  $
between local $t$-matrices of the physical and a reference
system. In the screened KKR method\cite{SKKR}
we use a lattice of strongly repulsive,
constant muffin-tin potentials (typically, $\sim 4$Ry height)
as reference system that leads to structure
constants which decay exponentially in real space.

\begin{figure}[t]
\begin{center}
\includegraphics[scale = 0.9]{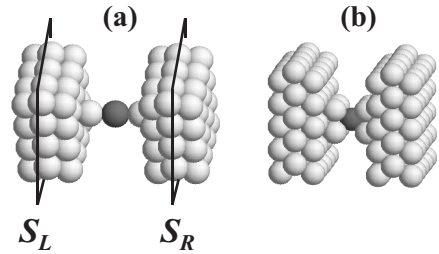}
\caption{Geometry of Cu and Pd atomic-sized constrictions
studied in the paper: (a) three atom chain suspended between the leads;
(b) pyramid-like contact. The central (contact) atom of the constriction
can be either a host or impurity atom. The conductance is calculated between
left ($S_{\mathrm{L}}$) and right ($S_{\mathrm{R}}$)
planes positioned in the leads.}
\end{center}
\end{figure}

  When the KKR method is applied to the systems shown in Fig.1 both the
constriction region and the leads are treated on the same footing.
This is achieved by using the hierarchy of Green's functions
connected by a Dyson equation, so that one performs
the self-consistent electronic structure
calculations of complicated systems in a step-like manner.
First, using the concept of principal layers
together with the decimation technique,\cite{2D_SKKR}
we calculate the structural Green's function
${G}^{0\,jj'}_{LL'}({\bf k}_{\parallel},E)$
of the auxiliary system consisting of semi-infinite leads
separated by a vacuum barrier
(here $j$ and $j{\,'}$ refer to layer indexes).
At the the second step the atomic cluster is embedded between the
leads by solving the Dyson equation self-consistently,
\begin{eqnarray}
\label{Dyson}
G^{nn'}_{LL'}(E) & = & G^{0\,nn'}_{LL'}(E)  \\
\nonumber
 & & +  \sum_{n''L'' } G^{0\,nn''}_{LL''}(E)\,
\Delta t^{n''}_{L''}(E)\, G^{n''n'}_{L''L'}(E),
\end{eqnarray}
and the structural Green's function $G^{nn'}_{LL'}(E)$
of the complete system is obtained.
Eq.~(\ref{Dyson}) is solved in real space since
due to effective screening of the perturbation
the charge deviations are restricted nearby the constriction.

\subsection{Evaluation of conductance}

 In what follows, we will consider only a ballistic coherent regime of the electron
transport thus limiting our study to zero temperature, infinitely small
bias and singe-particle picture. To evaluate the conductance of nanocontacts
we employ the Kubo linear response theory as formulated by Baranger
and Stone: \cite{BarStone}
\begin{eqnarray}
\nonumber
g & = & G_0
\int_{S_{\mathrm{L}}}dS\int_{S_{\mathrm{R}}}dS\,' \\
 & & \times\ G^{+}({\bf r},{\bf r'},E_F)
\stackrel{\leftrightarrow}\partial_z
\stackrel{\leftrightarrow}\partial_{z'} G^{-}({\bf r'},{\bf r},E_F),
\label{cond_g}
\end{eqnarray}
here $G_0 = 2e^2/h$, $G^{-}$ and $G^{+}$ are retarded and advanced
Green's functions, respectively. The current flows in
$z$ direction, and $f\stackrel{\leftrightarrow}\partial_zg =
f (\partial_z g) - (\partial_z f)g$.
The integration is performed over two (left and right)
planes $S_{\mathrm{L}}$ and $S_{\mathrm{R}}$ which connect
the leads with the scattering region as it is shown in Fig.~1.
The implementation of the Eq.~(\ref{cond_g}) in the
site angular momentum representation of the KKR
method and related to it convergence properties
were described in detail in Ref.~\onlinecite{Mavropoulos}.
Within the ASA used in this work instead of integration over planes
we average conductance over atomic layers and obtain:
\begin{equation}
\label{g_TraceDGDG}
g\ =\ G_0\, \mathrm{Tr}_{(n,L)}
\left[D_{\mathrm{L}}\, G\, D_{\mathrm{R}}\, G^{\dagger}
\right].
\end{equation}
Here, Tr involves site ($n$) and angular momentum ($L$) indices
related to the atomic plane~$S_{\mathrm{L}}$ (Fig.~1).
$G  = \{ G^{nn'}_{LL}(E_F) \} $
stands for the matrix notation of the structural Green's function
introduced in Eq.~(\ref{Grr}), taken at the Fermi energy.
Site-diagonal operators $D_{\mathrm{L}}$
and $D_{\mathrm{R}}$ related to the left (L) and right
(R) leads are defined as
\begin{eqnarray}
\nonumber
\lefteqn{
\bigl[D_{\mathrm{L(R)}}\bigr]^{nn'}_{LL'}} \\
\label{D_LR}
& = & \pm\,
\frac{\delta_{nn'}}{\Delta}
\int\limits_{V^n_{\mathrm{L(R)}}} d^3r \left [R^{n\, *}_L({\bf r},E_F)\,
i\stackrel{\leftrightarrow}{\partial_z}
R^{n}_{L'}({\bf r},E_F) \right],
\end{eqnarray}
where different sings, "$+$" and "$-$",
refer to different operators, $D_{\mathrm{L}}$ and $D_{\mathrm{R}}$,
respectively. The integral is performed over the
volume $V^n_{\mathrm{L(R)}}$ of the Wigner-Seitz cell
around site $n$ which belongs either to the atomic plane $S_{\mathrm{L}}$ or
to the plane $S_{\mathrm{R}}$,
and $\Delta$ is the distance separating nearest-neighbor
atomic layers in $z$ direction.

\subsection{Conductance eigenchannels}

Ballistic conductance of an atomic constriction can be decomposed into
individual eigenchannels. According to the Landauer formalism,
which was proved to be equivalent to the Kubo approach,\cite{BarStone}
conductance reads as $g = G_0 \sum_n \tau_n(E_F)$. The $\tau_n$'s are
eigenvalues of the matrix $T_{\mathbf{k}\mu; \mathbf{k'}\mu'}$ which
defines the transmission probability for an
incident Bloch wave from the left electrode
to be transmitted to an outgoing Bloch wave in the right, where
$\mathbf{k}$ and $\mathbf{k'}$ are Bloch vectors for incident
and transmitted waves, and $\mu$ and $\mu'$ are the corresponding band indices.
A procedure for the evaluation of the conductance eigenchannels will be published
elsewhere.\cite{KKR_channels}
To solve the eigenvalue problem for the transmission
matrix $T$, the perturbed Bloch states of the whole system
are projected to the basis of local functions
$R_L^n(\mathbf{r})$ and conductance
is represented as a sum of two contributions:
\begin{equation}
\label{g_2terms}
g\ =\  G_0\, \mathrm{Tr\,}_{(n,L)}
\left[{\cal{T}}^{+} \right]\ +\
G_0\, \mathrm{Tr\,}_{(n,L)} \left[{\cal{T}}^{-} \right],
\end{equation}
with
\begin{equation}
{\cal{T}}^{\pm} = \pm
\sqrt{\pm D^{\pm}_{\mathrm{L}}}\,
G\, D_{\mathrm{R}}\, G^{\,\dagger}
\sqrt{\pm D^{\pm}_{\mathrm{L}}}.
\end{equation}
Here $D^{\pm}_{\mathrm{L}}$ are two parts of the anti-symmetric hermitian
operator $D_{\mathrm{L}} =  D^{+}_{\mathrm{L}} + D^{-}_{\mathrm{L}}$
defined by Eq.(\ref{D_LR}).
The $D^{+}_{\mathrm{L}}$ and $D^{-}_{\mathrm{L}}$ have
positive and negative eigenvalues, respectively.

When the atomic plane $S_{\mathrm{L}}$ is placed in the asymptotic region of the left
lead far from the atomic constriction the second term in
Eq.(\ref{g_2terms}), involving trace of ${\cal T}^{-}$,
is equal to zero and eigenchannels are found as eigenvectors
of operator ${\cal T}^{+}$.
However, in practice we are forced to take integration planes
closer to the constriction in order to obtain convergent value for the conductance
with respect to number of atoms in the cross-sections of the leads.
The positions of the atomic planes $S_{\mathrm{L}}$ and
$S_{\mathrm{R}}$, therefore,
do not meet the asymptotic limit criterium as it is formally required
by the scattering approach. However, since the current
through the structure is conserved, any position of the
planes is suitable for the calculation of conductance.
Thus, if $S_{\mathrm{L}}$ is placed somewhere in the scattering
region we have to sum up all multiple scattering contributions.
All contributions in direction of the current
cause ${\cal T}^{+}$, whereas all multiple scattering contributions
in opposite direction of the current give rise to ${\cal T}^{-}$.
In the region of the lead where the potential is a small perturbation
with respect to the bulk potential the contribution to the conductance
due to ${\cal T}^{-}$ is one order of magnitude smaller than ${\cal T}^{+}$.
To identify the transmission probabilities of individual
eigenchannels the spectrums of the operators
${\cal T}^{+}$ and ${\cal T}^{-}$ are arranged in a proper way
using the symmetry analysis of eigenvectors.
For further details we refer to Ref.~\onlinecite{KKR_channels}.

Most of the structures studied in this work obey the
$C_{4v}$ symmetry determined by the fcc (001) surface (Fig.1).
Further we denote individual channels by the indices
of the irreducible representations of this group
using notations of Ref.~\onlinecite{Wigner}, common in the band
theory. In addition, each channel can be classified
according to the angular momentum contributions
when the wave function of the channel is projected
on the contact atom of the constriction. This is very
helpful since the channel transmission can be related
to the states of the contact atom.\cite{Cuevas_TB_model}
For example, the identity representation $\Delta_1$
of the $C_{4v}$ group is compatible with the  $s$, $p_z$ and $d_{z^2}$ orbitals
(here $z$ is the axis perpendicular to the surface),
while the the two-dimensional representation $\Delta_{5}$
is compatible with the $p_x$, $p_y$, $d_{xz}$,  $d_{yz}$ orbitals.

\begin{figure}[t]
\begin{center}
\includegraphics[scale = 1.05]{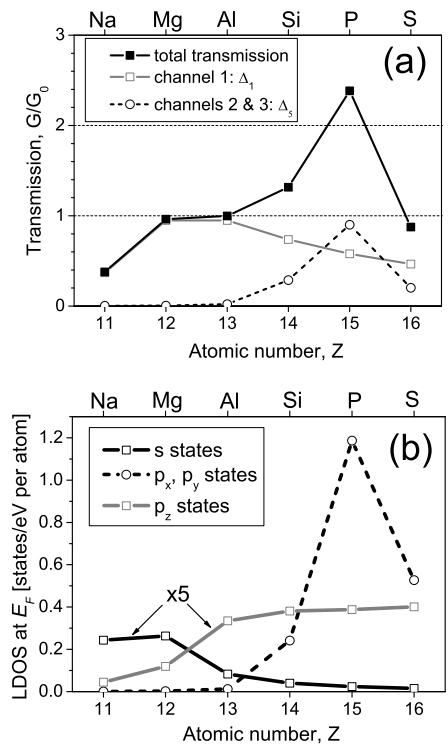}
\caption{Top panel (a): conductance and corresponding eigenchannel decomposition
through the atomic chains with $sp$ impurities connected to the Cu leads
[see Fig.1(a)]. Bottom panel (b): Symmetry resolved LDOS
at impurity sites, at the Fermi energy, {\it vs} atomic number of impurity.}
\end{center}
\end{figure}

\section{Results and discussion}

\subsection{Conductance through 
$\boldsymbol{sp}$ atoms}

We consider first atomic constrictions shown in Fig.1a which are
simulated by the straight three atom chains suspended between
copper leads with different impurities in the center of the junction.
The calculated conductance value of a pure Cu nanocontact is
$0.96\ G_0$. Only one channel contributes to the total transmission,
it has $\Delta_1$ symmetry and is projected
on $s$ and $p_z$ orbitals of the contact Cu atom.
In addition, small contributions from states with higher angular momentum
($d_{z^2}$ and $f_{z^3}$) are also present. This result
is in agreement with other theoretical studies\cite{TB_models,Ab-initio_1} as well as
with experiments for noble metals.\cite{Ludoph,Yanson_PhD}

  We proceed with the discussion of the conducting properties of single
$sp$ atoms ($Z = 11 \dots 16$) attached to Cu electrodes (Fig.1a).
In Fig.2 (top panel) we present the conductance
together with the eigenchannel decomposition and compare it
with the symmetry projected local density of states (LDOS)
at the impurity atom, at the Fermi energy (bottom panel).
We observe from Fig.2a that the conductance changes
significantly depending on the type of $sp$ impurity bridging
the Cu leads. For example, conductance through
the sodium atom is smaller than $0.5\,G_0$ while in case of phosphorus
it reaches a value of $2.4\,G_0$.
Our calculations show that three channels are involved in conductance:
one $sp_z$-channel of $\Delta_1$ symmetry
and a double degenerate channel of $\Delta_5$
symmetry which locally consists of $p_x$, $p_y$ states
when projection on the contact atom is performed.
From Fig.2 we see that the variation of the eigenchannels transmission
with atomic number follows the changes in the
orbital resolved LDOS at $E_F$.

\begin{figure}[t]
\begin{center}
\includegraphics[scale = 1.00]{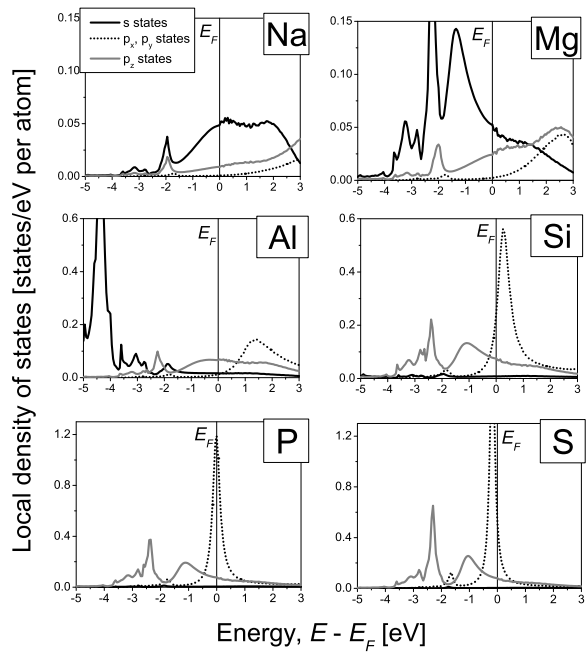}
\caption{Symmetry resolved LDOS at impurity atoms
placed at the contact site of Cu junction [see Fig.1(a)].}
\end{center}
\end{figure}

In case of Na, Mg and Al the transmission is essentially determined by
one highly symmetric $\Delta_1$ channel (Fig.2a).
For sodium the electronic states at $E_F$ have mainly $s$ character (Fig.3)
but transmission is small because of weak coupling to the Cu leads.
As we move up to Mg and Al the $s$ states are filled up with more electrons.
At the same time, the $p_z$ states contribute significantly to
the LDOS at $E_F$ (Fig.3). That causes a saturation of the
$\Delta_1$ channel transmission. For larger atomic numbers
the $\Delta_1$ channel is determined by $p_z$ states only.
The transmission of the $\Delta_1$ channel is reduced again since
for Si, P and S impurities the $s$ states are occupied.
The results of the calculations shown in Figs.~2 and 3 reveal that
the increase of conductance through Si, P and S impurities is due to a
virtual bound state formed by $p_x$ and $p_y$ orbitals (dashed line in Fig.3).
This state gives rise to a double degenerate $\Delta_5$ channel.
The $p_xp_y$ resonance, which is situated
just above $E_F$ for Si, is pinned to the Fermi level
in case of P, therefore the $\Delta_5$ channel becomes highly open.
For the S impurity the virtual bound
state is occupied and the transmission of the $\Delta_5$ channel drops.

\begin{figure}[t]
\begin{center}
\includegraphics[scale = 0.85]{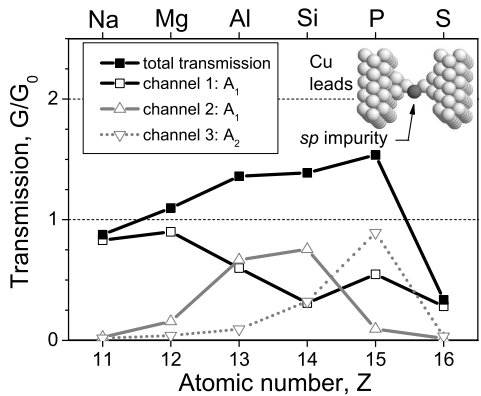}
\caption{Conductance and its eigenchannel decomposition for the
ziz-zag atomic chains suspended between the Cu leads with $sp$ atoms
placed at the contact site.}
\end{center}
\end{figure}

\begin{figure}[t]
\begin{center}
\includegraphics[scale = 1.10]{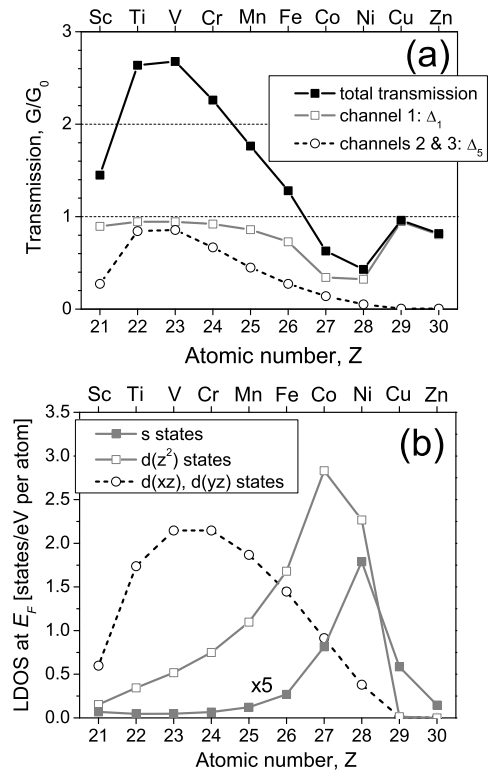}
\caption{Top panel (a): conductance and corresponding eigenchannel decomposition
through the atomic chains with $3d$ impurities connected to the Cu leads
[see Fig.1(a)]. Bottom panel (b): Symmetry resolved LDOS
at impurity sites, at the Fermi energy, {\it vs} atomic number of impurity.}
\end{center}
\end{figure}

To investigate the effect of atomic rearrangements
on conductance we moved the $sp$ atom from its symmetric position.
The nanocontact is now simulated by a zig-zag-like chain
suspended between Cu leads where the chain is
a continuation of the fcc structure along the [001] direction
(we show the exact geometry in the insert of Fig.4).
The behavior of the conductance {\it vs} the atomic number (Fig.4)
is similar to the previous case:
transmission is smaller than 1.0 for Na, has a maximum for P and drops
down for S. However, the absolute values are different for
the two considered configurations. We found that the conductance
consists again of three channels. Since the symmetry is reduced
all channels are different and the degeneracy is lost.
The 1st and 2nd channels correspond to an identity
representation ($A_g$) of
the $S_2$ group, while the 3rd channel is an antisymmetric one  ($A_u$).

\subsection{Conductance through 
3$\boldsymbol{d}$ atoms attached to Cu leads}

As we have seen, when atomic orbitals are present at
the Fermi level they support additional conducting channels.
That should be also true for the case of $3d$ transition
metal impurities. They have five localized $d$ states
nearby $E_F$ which potentially might be involved
in the formation of new channels and
the conductance could change as compared with
$\approx 1.0\,G_0$ found for a Cu constriction.
Our results for the conductance of the systems with 3$d$ transition
metal impurities are given in Fig.5. We consider a single atom
contact modelled by a short atomic chain
connecting the Cu leads, with impurity inside the chain
(Fig.1a). Indeed, we see that conductance varies
in a broad range along the 3$d$ series reaching
values of about $2.7G_0$ for Ti and V, and decreasing
to $0.5G_0$ in the case of a Ni atom.
We have excluded many body effects (e.g., Kondo effect)
from the consideration. As an approximation in the calculations the 3$d$ impurities
were supposed to have zero magnetic moments. That could be
justified by the fact that at $T=0$ the magnetic moment
of the impurity is effectively screened by the conduction electrons.

\begin{figure}[t]
\begin{center}
\includegraphics[scale = 1.10]{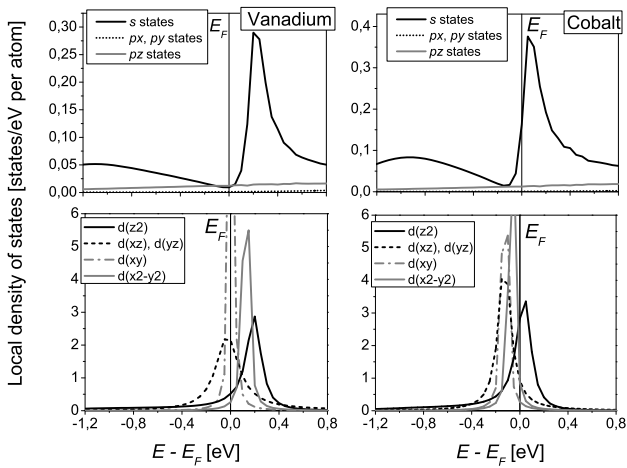}
\caption{Symmetry resolved LDOS at impurity atoms
placed at the contact site of Cu junction [see Fig.1(a)].}
\end{center}
\end{figure}

\begin{figure}[b]
\begin{center}
\includegraphics[scale = 1.00]{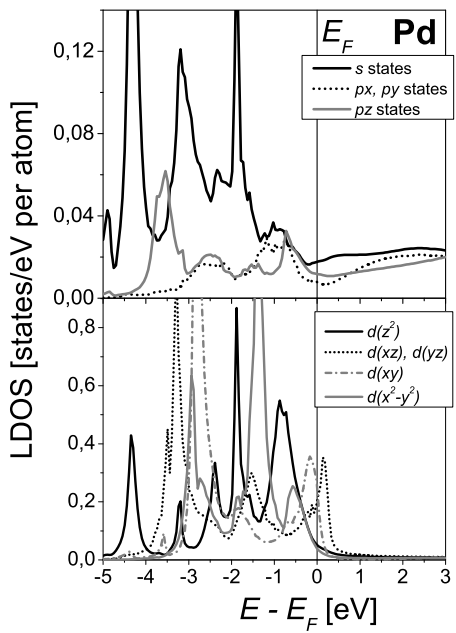}
\caption{Symmetry resolved LDOS at the central atom of the Pd junction
shown in Fig.1(b).}
\end{center}
\end{figure}

Conductance is found to be a sum of three channels with
$\Delta_1$ and $\Delta_5$ symmetry, where the last
channel is two fold degenerate.
To understand the behavior of channels transmission
(Fig.5a) we present in Fig.5b the symmetry
projected LDOS at the $3d$ impurities at the Fermi energy,
as well as in the vicinity of $E_F$, for selected atoms (Fig.6).
Calculations show that the LDOS nearby $E_F$ is formed due to $4s$ and $3d$
states of transition metal atoms while $4p$ states are above $E_F$ and thus
play no role for transport. As for $3d$ states,
they form sharp resonances localized in the energy
window of 1~eV around the Fermi level (Fig.6, low panels).
The $d_{xy}$ and $d_{x^2-y^2}$ resonance states,
even if they are situated at the Fermi level,
do not support open channels. Since
the $d_{xy}$ and $d_{x^2-y^2}$ orbitals are localized perpendicular
to the wire axis effective coupling to the neighboring atoms is prevented.
Consequently, the corresponding resonances look like
sharp peaks in the LDOS (Fig.6, down panels).

Consider first the $\Delta_5$ channel which is projected on $d_{xz}$, $d_{yz}$
states. The two-fold degenerate resonance state of this symmetry
is situated above $E_F$ in case of Sc that leads
to a relatively small transmission of the corresponding channel.
When more $d$ electrons are present at the contact atom,
the $d$ states move to lower energies. In particular,
for Ti, V and Cr the $d_{xz},d_{yz}$ impurity state
is situated very close to the Fermi level (Fig.6, down left panel)
and causes high transmission of the $\Delta_5$ channel.
For larger atomic number the $d_{xz},d_{yz}$ resonance
moves to energies below $E_F$ and transmission
of the corresponding channel is reduced.

The highly symmetric $\Delta_1$ channel is open and is mainly
due to $s$ states for the first half of the $3d$ series, from Sc to Mn.
The increase of the $s$ and $d_{z^2}$ LDOS in case of Fe, Co and Ni
(Fig.5b) is accompanied by a reduction of the transmission.
That is, because of the hybridization between $s$ states
and the $d_{z^2}$ resonance (Fig.6, down right panel)
with $\Delta_1$ symmetry, the wave function becomes more localized
at the impurity atom. From another point of view,
the $\Delta_1$ channel is scattered at the $d_{z^2}$ impurity state.
For Cu and Zn the $d$ states are shifted to energies
far below $E_F$, so that only a single $s$-channel is left.

\begin{figure}[b]
\begin{center}
\includegraphics[scale = 1.00]{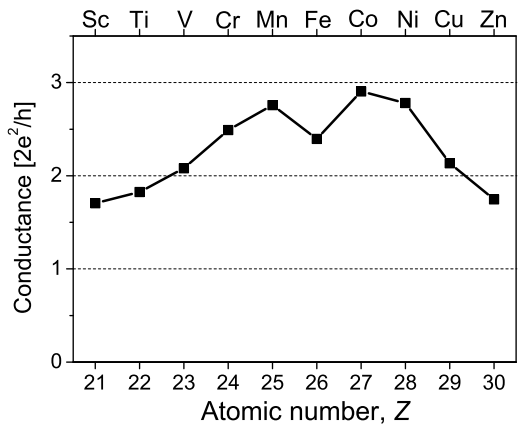}
\caption{Conductance through Pd pyramid-like junction
contaminated by $3d$ impurities [see Fig.1(b)].}
\end{center}
\end{figure}

\begin{figure*}
   \begin{minipage}[lc]{0.75\textwidth}
    \centering \includegraphics[scale = 1.60]{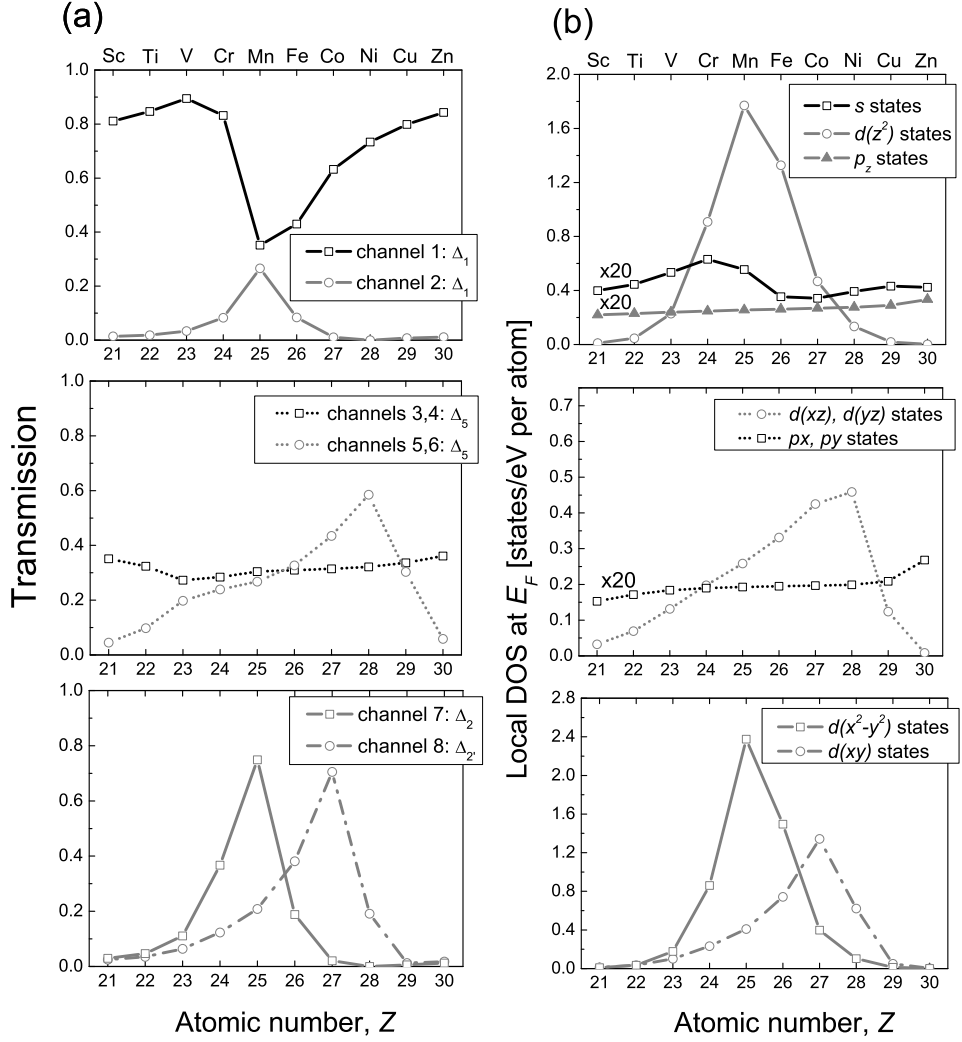}
   \end{minipage}
   \begin{minipage}[c]{0.24\textwidth}
\caption{(a) Left column: transmission probabilities of eigenchannels
through Pd junctions contaminated by $3d$ impurities; (b) Right column: symmetry resolved
LDOS at impurity atoms.}
   \end{minipage}
\end{figure*}

\subsection{Pd constrictions with transition metal impurities}

One could expect a variety of different eigenchannels for atomic
constrictions made of transition metals.
An example of such a system is a Pd pyramid like contact shown in Fig.1(b),
with conductance being $2.8\,G_0$.
The LDOS at the contact atom is mainly $d$-like nearby the Fermi energy,
as it is seen from Fig.7. This is also the case for the Pd surface.
Since $d$ states of different symmetry are available at the Fermi
energy (Fig.7), six channels give a main contribution to conductance. These are
one channel of $\Delta_1$ symmetry: $\tau_1 = 0.72$ ($sp_z$-like);
two double generate channels of $\Delta_5$ symmetry:
$\tau_2 = \tau_{2'} = 0.58$ ($d_{xz},d_{yz}$-like) and
$\tau_3 = \tau_{3'} = 0.32$ ($p_{x},p_{y}$-like);
and one channel of $\Delta_{2'}$ symmetry, $\tau_4 = 0.24$ ($d_{xy}$-like).
We would like to mention, that the channel with the highest symmetry
($\Delta_1$) has the largest transmission. Here,
the irreducible representations belong to the $C_{4v}$ group.

 Let us now study a situation where the contact atom
of the otherwise Pd junction is substituted by $3d$
transition metal impurities. The conductance variation along
the $3d$ series is shown in Fig.8.
We observe that a Ni impurity does not affect conductance value
of $2.8\,G_0$ of a Pd contact simply because Ni is isovalent to Pd.
For other impurities conductance changes from $1.7\,G_0$
in case of Sc and Zn, up to $2.9\,G_0$ in the case of a Co atom.
Eight channels give rise to the total transmission.
Results of the individual transmission probabilities are presented
in Fig.9 (left column) together with the LDOS at $E_F$ projected
on the corresponding orbitals of the impurity atoms (right column).
Note, that due to a larger opening angle as compared
with the first atomic configuration shown in Fig.1,
the channels of $\Delta_{2'}$ ($d_{xy}$) and
$\Delta_2$ ($d_{x^2-y^2}$) symmetry
contribute significantly to conductance, while
they were closed in the case of the Cu constrictions with
$3d$ impurities discussed in the previous section.

Without showing a detailed structure of the LDOS nearby the Fermi energy,
we mention that a high density at $E_F$ of the $d$ states
of different symmetry is always related to resonance-like states
(see e.g.\ Fig.7 for the Pd atom).
When some of these resonances approach the Fermi level,
the transmission of the corresponding channel reaches a maximum.
Such a situation takes place for the $\Delta_5$ channel ($d_{xz},d_{yz}$-like)
in case of Ni (Fig.9, middle panels), as well as for the $\Delta_2$ ($d_{x^2-y^2}$)
and $\Delta_{2'}$ ($d_{xy}$) channels in case of Mn and Co, respectively
(Fig.9, lower panels). For pure $d$ channels mentioned above,
a clear correlation between the partial LDOS at $E_F$
and the transmission probabilities is observed.
The other channel of $\Delta_5$ symmetry does not
show a strong variation of the transmission {\it vs}
the atomic number of the impurity. This channel is
mainly the $p_x,p_y$-like.

\begin{figure*}
 \begin{minipage}[lc]{0.75\textwidth}
    \centering \includegraphics[scale = 1.5]{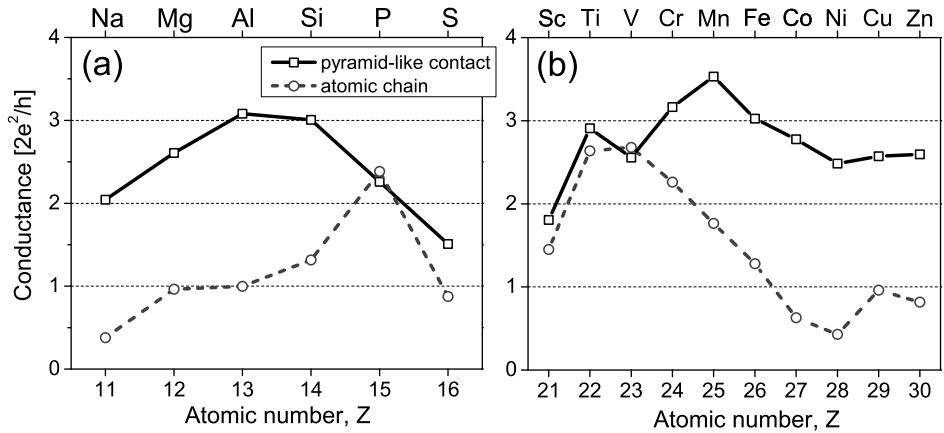}
 \end{minipage}
 \begin{minipage}[c]{0.24\textwidth}
   \caption{Conductance through $sp$ (left plot) and $3d$ (right plot)
   transition metal atoms attached
   to Cu leads in a different way as shown in Fig.1. Solid line:
   pyramid-like configuration (Fig.1b), dashed line: atomic chain (Fig.1a).
   The influence of the electrode material on the conductance
   can be seen if we compare solid line in plot (b) with results of Fig.8, where
   a geometry of the constriction was not changed but Pd was used instead of~Cu.}
\end{minipage}
\end{figure*}

Finally, we note that two different channels of $\Delta_1$
symmetry exhibit an interesting variation.
One of them is $sp_z$-like, it is open at the
beginning and at the end of the $3d$ series (Fig.9, upper panels).
The $d_{z^2}$ resonance approaching the Fermi level in case of Mn and
Fe impurities acts in two ways. Firstly, due to a hybridization with the $s$ state
it reduces the transmission of the already existing $sp_z$ channel
(that is similar to the case discussed in the previous section, Fig.5).
Secondly, the $d_{z^2}$ resonance itself opens
a new channel of $\Delta_1$ symmetry, which consequently is
the $d$-like. Its transmission follows the variation of the $d_{z^2}$-LDOS
when the atomic number is changed.

\subsection{Does an atom possess certain conductance?}

We have learnt from presented examples that conductance properties
of individual atoms can be analyzed in detail with help of
eigenchannel analysis. Therefore the question arises: can
we attribute definite conductance to the particular atom?
Comparing transmission through $3d$ transition metal impurities
attached to Cu and Pd leads (Fig.5a and Fig.8, respectively), we
see that the answer to this question is in general negative. This
statement is also confirmed by results presented in Fig.10, where
we compare transmission through $sp$ and $3d$ transition metal atoms
attached to Cu electrodes in a different way as it is shown in Fig.1.
We have mentioned in the previous section, that due to a larger opening
angle for the incident waves, conductance of pyramid-like
constrictions increases as compared to nanocontacts simulated by
atomic chains. The number of non-vanishing conducting channels
needed to describe the conductance along $sp$ and $3d$ series becomes
also larger. Without going deep into details, we point out that
number of eigenmodes increases up to 4 in case of $sp$ impurities
(Fig.10a) and up to 6 in case $3d$ transition metal atoms
(Fig.10b). In particular, in the latter case,
because of the increased coordination number of the contact atom,
the $d$ channels of $\Delta_{2'}$ ($d_{xy}$) and
$\Delta_{2}$ ($d_{x^2-y^2}$) symmetry become open for Ti and Mn, respectively.
These channels were closed in the case of a nanocontact
modelled by a three atom chain (see section III.B for discussion).
Thus, the presented analysis gives no evidence to
attribute a precise value of conductance to the given atom.
A conclusion is that the conductance is sensitive to the geometry of
atomic constriction as well as to the type of electrodes.
However, our results show that for the specified single-atomic contact
eigenchannels can always be related to the
electronic states of the appropriate symmetry at the Fermi energy.

\section{Summary}

In conclusion, based on the KKR Green's function method combined
with the Kubo approach we have performed parameter-free {\it ab initio}
calculations of the ballistic conductance through single
$sp$ and $3d$ transition metal atoms attached
to Cu and Pd leads. We have investigated in which way the valency of the atom
bridging two metallic leads is responsible for the number
of conducting channels. We have found that both the
chemical valency of the atom and the opening angle of
the atomic constriction determine the number
of eigenmodes. In accordance with the known tight-binding model
by Cuevas {\it et al.},\cite{Cuevas_TB_model} we have confirmed by our study
that the symmetry of the open conducting channels is related
to the symmetry of the electronic states at the contact site.
That allows to relate the open channels
to the orbitals of the contact atom available at the Fermi energy.
Furthermore, we have shown that impurity resonances
approaching the Fermi level can open new channels of the appropriate symmetry.
On the other hand, in some cases, the transmission of the existing channels
can even be blocked by scattering at virtual
bound states which have in this case the symmetry compatible
with the symmetry of the open channels.

\section*{Acknowelegment}

This work was supported by the Deutsche Forschungsgemeinschaft (DFG),
Schwerpunktprogramm 1165 "Nanodr{\"a}hte und Nanor{\"o}hren".


\end{document}